\documentclass[draft,grl]{agutex}

\authorrunninghead{YERMOLAEV}

\titlerunninghead{COMMENT}

\begin{document}

%
%

\title{Comment on "Are periodic solar wind number density structures formed in the solar corona?" \\ 
by N. M. Viall  et al.}
%

%
%



\authors{Yu. I. Yermolaev \altaffilmark{1}
}

\altaffiltext{1}{Space Plasma Physics Department, Space Research Institute, 
Russian Academy of Sciences, Profsoyuznaya 84/32, Moscow 117997, Russia. 
(yermol@iki.rssi.ru)}






%
%


\begin{abstract}

Location of formation of periodic solar wind number density structures is discussed. 
Observation of proton and alpha anticorrelation in these structures 
\citep{Vialletal2009} indicates that 
taking into account that bulk velocity of aplha-particles is higher than that of proton 
the place of formation for these structures should be located at distance less 0.002 AU  
from place of observation.  

\end{abstract}

%
%

%

\begin{article}

%
%

The periodic density fluctuations
with period $\sim$~10~min are observed sometimes in the solar wind at 1 AU 
\citep{Kepkoetal2002,KepkoSpence2003}.
Recently 
\cite{Vialletal2009} 
analyzed WIND observations of variations in ratio of alpha to proton density (AHe) during an event on 14 February, 1996, containing periodic ($\sim$~30 min) proton density fluctuations in order to determine whether the periodic density fluctuations developed in the interplanetary medium, or if they were instead more likely generated somewhere within the solar corona. As the proton density fluctuates in a periodic nature, the alpha density also fluctuates at the same periodicity in antiphase with the protons. On the basis of these observations the authors conclude that "For this event, the anti-phase nature of the AHe variations strongly suggests that periodic solar wind density structures originate in the solar corona." In the present comment we will show that these conclusions are based on incorrect assumptions and the presented data say in favor of a local origin of periodic structure.

\cite{Vialletal2009} 
implicitly assume that bulk velocities of protons and alpha-particles are equal during their motion from the Sun to the Earth. As have been shown by numerical space experiments the alpha-particle velocity is typically larger than that of protons and difference of their velocities changes from low value (about 0) in very slow (Heliospheric Current Sheet) streams up to comparable to but less than the Alfven velocity in fast streams of solar wind 
\citep{Formisanoetal1970,Robbinsetal1970,Bolleaetal1972,Ogilvie1975,Asbridgeetal1976,Bosquedetal1977,Neugebauer1981,Marschetal1982,Ogilvieetal1982,YermolaevandStupin1997,Yamauchietal2004}.
At solar wind velocity of 500 km/s at 1 AU the alpha-particle velocity is higher about (2-5)~\% than proton one. As a result both components which have been together when they have left the Sun in the interplanetary space, can have spatial shift at 1~AU rather each other equal (0.02 - 0.05) AU or temporal shift of (100 - 250) min. As the difference in component velocities grows at approach to the Sun 
\citep{Marschetal1982,YermolaevandStupin1990}  
real shift is larger. This shift is significantly larger than period of structure ($\sim$~30 min). It means that anticorrelation of density of protons and alpha-particles, which was formed at the Sun, is completely absent near the Earth because two components are shifted in space  rather each other on distances which exceed both the structure period, and the size of the phenomenon. 

Thus, the observed anti-correlation of densities of protons and alpha particles can be born near an observation point (no more than 0.002 AU) that the shift between components due to their velocity difference was less period of variations.



%
%
%
%
%
%

%
%
%
%

\begin{acknowledgments}
The work was support by Russian Foundation of Basic Research, grant 10-02-00277.
\end{acknowledgments}

%
%
%
%
%
%
%
%
%
%




%
%

\end{article}




%
%
%
%
%
%


\end{document}